\documentclass[a4paper,twoside,onecolumn]{IEEEtran}
\usepackage{amsmath}
\usepackage{amssymb}
\usepackage{xcolor}

\newtheorem{theo}{Theorem}
\newtheorem{lemm}[theo]{Lemma}
\newtheorem{coro}[theo]{Corollary}
\newtheorem{claim}[theo]{Claim}

\newcommand{\calW}{\mathcal{W}}
\newcommand{\calX}{\mathcal{X}}
\newcommand{\calY}{\mathcal{Y}}
\newcommand{\calZ}{\mathcal{Z}}

\newcommand{\bfp}{\mathbf{p}}

\newcommand{\bfu}{\mathbf{u}}

\newcommand{\Vol}{\mathrm{Vol}}

\newcommand{\mysett}[1]{\{#1\}}
\newcommand{\mysizee}[1]{|#1|}

\newcommand{\reals}{\mathbb{R}}
\newcommand{\lambdamin}{\lambda_{\mathrm{min}}}

\newcommand{\normts}[1]{\left\| #1 \right\|_2^2} 
\newcommand{\normt}[1]{\left\| #1 \right\|_2} 

\newcommand{\degradingCost}{\mathrm{DC}}
\newcommand{\degraded}{\prec}

\newcommand{\Deltatilde}{\tilde{\Delta}}

\newcommand{\Whard}{\mathsf{W}}
\newcommand{\sphere}{\mathrm{sphere}}

\newcommand{\probvectorr}[1]{( #1 )}
\newcommand{\probvector}[1]{\left( #1 \right)}
\newcommand{\probvectorm}[1]{\langle #1 \rangle}

\newcommand{\outputAlphabet}[1]{\mathrm{out}(#1)}
\newcommand{\channelSize}[1]{\mysizee{\mathrm{out}(#1)}}

\newenvironment{condmultline}{\begin{equation}}{\end{equation}}

\newenvironment{condmultlinestar}{\begin{equation*}}{\end{equation*}}

\newcommand{\bibfilePath}{../../../../ee/mybib}
\newcommand{\twobibs}[2]{#2} 

\title{On the Construction of Polar Codes for Channels with Moderate Input Alphabet Sizes}
\author{{Ido Tal}\\
\authorblockA{
Department of Electrical Engineering,\\
Technion, Haifa 32000, Israel.\\
Email: {\tt idotal@ee.technion.ac.il}
}
\thanks{The paper was presented in part at the
2015 IEEE International Symposium on Information Theory,
Hong Kong, June 14 -- June 19, 2015.
Research supported in part by the Israel Science Foundation
grant 1769/13.}
}

\begin{document}
\maketitle
\begin{abstract}
Current deterministic algorithms for the construction of polar codes can only  be argued to be practical for channels with small input alphabet sizes. In this paper, we show that any construction algorithm for channels with moderate input alphabet size which follows the paradigm of ``degrading after each polarization step'' will inherently be impractical with respect to a certain ``hard'' underlying channel. This result also sheds light on why the construction of LDPC codes using density evolution is impractical for channels with moderate sized input alphabets.
\end{abstract}

\section{Introduction}
Polar codes \cite{Arikan:09p} are a novel family of error correcting codes which are capacity achieving and have efficient encoding and decoding algorithms. Originally defined for channels with binary input, they were soon generalized to channels with arbitrary input alphabets \cite{STA:09a}. Although polar codes are applicable to many information theoretic settings, the channel coding setting is the one we consider in this paper. More specifically, we consider the symmetric capacity setting discussed in \cite{Arikan:09p} and \cite{STA:09a}.
In this setting, a polar code is gotten by unfreezing channels with probability of error at most $2^{-\beta n}$, where $n$ is the code length and $\beta>0$ is a suitably chosen constant. A synthesized channel is gotten by repeatedly applying polar channel transforms. The plus and minus polar transforms were defined in \cite{Arikan:09p}. Other transforms are possible \cite{MoriTanaka:10c}, \cite{MoriTanaka:14p} \cite{PresmanShapiraLitsyn:11c}, see also \cite{Sasoglu:12c}. 

Since the synthesized channels have an output alphabet size which grows exponentially in the code length $n$, calculating their probability of misdecoding is intractable if approached directly. To the author's knowledge, the only tunable and deterministic methods of circumventing this difficulty involve approximating some of the intermediate channels by channels which have a manageable output alphabet size. Simply put: before the first polarization step and after each polarization step, approximate the relevant channel by another channel having a prescribed output alphabet size. Doing so ensures that the channel output alphabet sizes do not grow intractably.

The above ``approximate after each polarization step'' idea has its origins in density evolution \cite[Page 217]{RichardsonUrbanke:08b}, a method to evaluate the performance of LDPC code ensembles. Density evolution was suggested as a method of constructing polar codes in \cite{MoriTanaka:09c}. In order to bound the misdecoding probability of a synthesized channel  --- as opposed to only approximating it --- one can force the approximating channel to be either (stochastically) degraded or upgraded with respect to it. An efficient algorithm for such a degrading/upgrading approximation was introduced for the binary-input case in \cite{TalVardy:13p} and analyzed in \cite{PHTT:11c}. See also \cite{KurkoskiYagi:11a} for an optimal degrading algorithm. Algorithms for degrading and upgrading non-binary channels were given in \cite{TalSharovVardy:12c} and \cite{PeregTal:15a}, respectively. See also \cite{GhayooriGulliver:12a}. On a related note, the construction of polar codes was recently proven to be polynomial \cite{GuruswamiVelingker:14a}, for an arbitrary but \emph{fixed} input alphabet size.

For a fixed input distribution, a degrading approximation results in a channel with reduced mutual information between input and output. This drop in mutual information should ideally be kept small. The reason for this will be elaborated on in Section~\ref{sec:codeImplications}. In brief, the reason is that such a drop necessarily translates into a drop in code rate, both in the polar coding setting as well as in the LDPC setting. Thus, a non-negligible drop in mutual information due to approximation necessarily means a coding scheme which is not capacity achieving. 

In this paper, we define a specific channel. With respect to this channel, we derive lower bounds on the drop in mutual information as a function of the channel input alphabet size, $q$, and the number of output letters of the approximating channel, $L$. Simply put, the main result of this paper is that for moderate values of $q$, a modest drop in mutual information translates into the requirement that $L$ be unreasonably large, in the general case. It seems to be common knowledge that constructing capacity achieving LDPC or polar codes for channels with such input alphabet sizes is generally hard; this is commonly referred to as the ``curse of dimensionality''. This paper is an attempt to quantify this hardness, under assumptions that are in line with what is currently done.

The structure of this paper is as follows. Section~\ref{sec:notation} introduces the main result of the paper, after stating the needed notation. Section~\ref{sec:codeImplications} explains the implications of the result to the hardness of constructing polar codes and LPDC codes. Section~\ref{Sec:preliminaryLemmas} contains a specialization of H\"older's defect formula to our setting. Section~\ref{sec:boundingTheDegradingCost} defines and analyzes the previously discussed channel.

\section{Notation and problem statement}
\label{sec:notation}
We denote a channel by $W \colon \calX \to \calY$. The probability of receiving $y \in \calY$ given that $x \in \calX$ was transmitted over $W$ is denoted $W(y|x)$. All our channels will be defined over a finite input alphabet $\calX$, with size $q = \mysizee{\calX}$. Unless specifically stated otherwise, all channels will have a finite output alphabet, denoted $\outputAlphabet{W} = \calY$. Thus, the channel output alphabet size is denoted $\channelSize{W}$.

We will eventually deal with a specific channel, which turns out to be symmetric (as defined in \cite[page 94]{Gallager:68b}). In addition, the input distribution we will ultimately assign to this channel turns out to be uniform.  However, we would like to be as general as possible wherever appropriate. Thus, unless specifically stated otherwise, we \emph{will not} assume that a generic channel $W$ is symmetric. Each channel will typically have a corresponding input distribution, denoted $P_X = P_X^{(W)}$. Note that $P_X$ \emph{need not} necessarily be uniform and \emph{need not} necessarily be the input distribution achieving the capacity of $W$. We denote the random variables corresponding to the input and output of $W$ by $X = X^{(W)}$ and $Y = Y^{(W)}$, respectively. The distribution of $Y$ is denoted $P_Y = P_Y^{(W)}$. That is, for $y \in \calY$,
\[
P_Y(y) = \sum_{x \in \calX} P_X(x) W(y|x) \; .
\]
The mutual information between $X$ and $Y$ is denoted as
\[
I(W) = I(X;Y) \; ,
\]
and is henceforth measured in nats. That is, all logarithms henceforth are natural. Note that $I(W)$ typically \emph{does not} equal the capacity of $W$.

We say that a channel $Q \colon \calX \to \calZ$ is (stochastically) degraded with respect to $W \colon \calX \to \calY$ if there exists a channel $\Phi \colon \calY \to \calZ$ such that the concatenation of $\Phi$ to $W$ yields $Q$. Namely, for all $x \in \calX$ and $z \in \calZ$,
\begin{equation}
\label{eq:degradingDefinition}
Q(z|x) = \sum_{y \in \calY} W(y|x) \Phi(z|y) \; .
\end{equation}
We denote $Q$ being degraded with respect to $W$ as $Q \degraded W$.

For input alphabet size $q = \mysizee{\calX}$ and specified output alphabet size $L$, define the \emph{degrading cost} as
\begin{equation}
\label{eq:degradingCostDefinition}
\degradingCost(q,L) \triangleq \sup_{W, P_X} \quad \min_{\substack{Q \; : \; Q \degraded W,\\ \channelSize{Q} \leq L}} \quad \left( I(W) - I(Q) \right) \; .
\end{equation}
Namely, both $W$ and $Q$ range over channels with input alphabet $\calX$ such that $\mysizee{\calX} = q$; both channels share the same input distribution $P_X$, which we optimize over; the channel $Q$ is degraded with respect to $W$; both channels have finite output alphabets and the size of the output alphabet of $Q$ is at most $L$; we calculate the drop in mutual information incurred by degrading $W$ to $Q$, for the ``hardest'' channel $W$, the ``hardest'' corresponding input distribution $P_X$, and the corresponding best approximation $Q$.

Note that the above explanation of (\ref{eq:degradingDefinition}) is a bit off, since the outer qualifier is ``$\sup$'', not ``$\max$''. Namely, we might need to consider a sequence of channels $W$ and input distributions $P_X$. Note however that the inner qualifier is a ``$\min$'', and not an ``$\inf$''. This is justified by the following claim, which is taken from \cite[Lemma 1]{KurkoskiYagi:11a}.
\begin{claim}
\label{cl:optimalQuantization}
Let $W \colon \calX \to \calY$ and $P_X$ be given. Let $L \geq 1$ be a specified integer for which $\mysizee{\calY} \geq L$. Then, 
\[
\inf_{\substack{Q \; : \; Q \degraded W,\\ \channelSize{Q} \leq L}}
\left( I(W) - I(Q) \right)
\]
is attained by a channel $Q \colon \calX \to \calZ$ for which it holds that $\channelSize{Q} = L$ and
\begin{condmultlinestar}
Q(z|x) = \sum_{y \in \calY} W(y|x) \Phi(z|y) \; , \qquad \Phi(z|y) \in \mysett{0,1} \; , \\ \qquad
\sum_{z \in \calZ} \Phi(z|y) = 1 \; .
\end{condmultlinestar}
Namely, $Q$ is gotten from $W$ by defining a partition $(A_i)_{i=1}^L$ of $\calY$
and mapping with probability $1$ all symbols in $A_i$ to a single symbol $z_i \in \calZ$, where $\calZ = \mysett{z_i}_{i=1}^L$.
\end{claim}

In \cite{TalSharovVardy:12c}, an upper bound on $\degradingCost(q,L)$ is derived. Specifically,
\[
\degradingCost(q,L) \leq 
2q\cdot \left( \frac{1}{L} \right)^{1/q}
\; .
\]
The above has been recently sharpened \cite[Lemma 8]{PeregTal:15a} to
\[
\degradingCost(q,L) \leq  2 \cdot q^{1 + \frac{2}{q-1}} \cdot \left( \frac{1}{L} \right)^{1/(q-1)}
\; .
\]

These bounds are constructive and stem from a specific quantizing algorithm. Specifically, the algorithm is given as input the channel $W$, the corresponding input distribution $P_X$, and an upper bound on the output alphabet size, $L$. Note that for a fixed input alphabet size $q$ and a target difference $\epsilon$ such that $\degradingCost(q,L) \leq \epsilon$, the above implies that we take $L$ proportional to $(1/\epsilon)^{q-1}$. That is, for moderate values of $q$, the required output alphabet size grows very rapidly in $1/\epsilon$. Because of this, \cite{TalSharovVardy:12c} explicitly states that the  algorithm can be considered practical only for small values of $q$. 

We now quote our main result: a lower bound on $\degradingCost(q,L)$. Let $\sigma_{q-1}$ 
be the constant for which the volume of a sphere in $\reals^{q-1}$ of radius $r$ is $\sigma_{q-1} r^{q-1}$. Namely,
\[
\sigma_{q-1} = \frac{\pi^{\frac{q-1}{2}}}{\Gamma(\frac{q-1}{2} + 1)} \; ,
\]
where $\Gamma$ is the Gamma function.
\begin{theo}
\label{theo:generalBound}
Let $q$ and $L$ be specified. Then,
\begin{condmultline}
\label{eq:generalBound}
\degradingCost(q,L) \geq \\ \frac{q-1}{2(q+1)} \cdot \left(\frac{1}{\sigma_{q-1} \cdot (q-1)!}\right)^\frac{2}{q-1}  
\cdot 
\left(\frac{1}{L} \right)^{\frac{2}{q-1}}
\; . 
\end{condmultline}
The above bound is attained in the limit for a sequence of symmetric channels, each have a corresponding input distribution which is uniform.
\end{theo}

The consequences of this theorem in the context of code construction will be elaborated on in the next section. However, one immediate consequence is a vindication of sorts for the algorithm presented in \cite{TalSharovVardy:12c}. That is, for $q$ fixed, we deduce from the theorem that the optimal degrading algorithm must take the output alphabet size $L$ at least proportional to $(1/\epsilon)^{(q-1)/2}$, where $\epsilon$ is the designed drop in mutual information. That is, the adverse effect of $L$ growing rapidly with $1/\epsilon$ is an inherent property of the problem, and is not the consequence of a poor implementation. For a numerical example, take $q=16$ and $\epsilon=10^{-5}$. The theorem states that the optimal degrading algorithm must allow for a target output alphabet size $L \approx 10^{23}$. This number is for all intents and purposes intractable.

We note that the term multiplying $(1/L)^{2/(q-1)}$ in (\ref{eq:generalBound}) can be simplified by Stirling's approximation. The result is that
\[
\degradingCost(q,L) \geq \quad \approx \frac{e}{4 \pi(q-1)}  \cdot \left(\frac{1}{L} \right)^{\frac{2}{q-1}} \; ,
\]
and the approximation becomes tight as $q$ increases.

Note that the RHS of the above is eventually decreasing in $q$, for $L$ fixed. However, it must be the case that $\degradingCost(q,L)$ is increasing in $q$ (to see this, note that the input distribution can give a probability of $0$ to some input symbols). Thus, we conclude that our bound is not tight.

\section{Implications for code construction}
\label{sec:codeImplications}
We now explain the relevance of our result to the construction of both polar codes and LDPC codes. In both cases, a ``hard'' underlying channel is used, with a corresponding input distribution that is uniform. Let us explain: for $q$ and $L$ fixed, and for a uniform input distribution, we say that a channel is hard if the drop in mutual information incurred by degrading it to a channel with at most $L$ output letters is, say, at least half of the RHS of (\ref{eq:generalBound}). Theorem~\ref{theo:generalBound} assures us that such hard channels exist. Put another way, the crucial point we will make use of is that for a hard channel, the drop in mutual information is at least proportional to $(1/L)^{2/(q-1)}$.

\subsection{Polar codes}
As explained in the introduction, the current methods of constructing polar codes for symmetric channels involve approximating the intermediate channels by channels with a manageable output alphabet size. Specifically, the underlying channel --- the channel over which the codeword is transmitted --- is approximated by degradation before any polarization operation is applied. Now, for $q$ fixed and $L$ a parameter, consider an underlying hard channel, as defined above. Denote the underlying channel as $W$, and let the result of the initial degrading approximation be denoted by $Q$.  

The key point to note is that the construction algorithm cannot distinguish between $W$ and $Q$. That is, consider two runs of the construction algorithm, one in which the underlying channel is $W$ and another in which the underlying channel is $Q$. In the first case, the initial degradation produces $Q$ from $W$. In the second case, the initial degradation simply returns $Q$, since the output alphabet size is at most $L$, and thus no reduction of output alphabet is needed. Thus, the rate of the code constructed cannot be greater than the symmetric capacity of $Q$, which is at most $W - \epsilon$. We can of course make $\epsilon$ arbitrarily small. However, this would necessitate an $L$ at least proportional to $(1/\epsilon)^{(q-1)/2}$. For rather modest values of $q$ and $\epsilon$, this is intractable. 

\subsection{LDPC codes}
The standard way of designing an LDPC code for a specified underlying channel is by applying the density evolution algorithm \cite[Section 4.4]{RichardsonUrbanke:08b}. To simplify to our needs, density evolution preforms a series of channel transformations on the underlying channel, which are a function of the degree distribution of the code ensemble considered. Exactly as in the polar coding setting, these transformations increase the output alphabet size to intractable sizes. Thus, in practice, the channels are approximated. If we assume that the approximation is degrading --- and it typically is --- the rest of the argument is now essentially a repetition of the argument above. In brief, consider an LDPC code designed for a hard channel $W$. After the first degrading operation, a channel $Q$ is gotten. The algorithm must produce the same result for both $W$ and $Q$ being the underlying channel. Thus, an ensemble with rate above that of the symmetric capacity of $Q$ will necessarily be reported as ``bad'' with respect to both $W$ and $Q$. Reducing the mutual information between $W$ and $Q$ is intractably costly for moderate parameter choices.

\section{Preliminary lemmas}
\label{Sec:preliminaryLemmas}
As a consequence of the data processing inequality, if $Q$ is degraded with respect to $W$, then $I(W) - I(Q) \geq 0$. In this section, we derive a tighter lower bound on the difference. To that end, let us first define $\eta(p)$ as
\[
\eta(p) = -p \cdot \ln p \; , \quad 0 \leq p \leq 1 \; ,
\]
where $\eta(0) = 0$. Next, for a probability vector $\bfp = \probvectorr{p_x}_{x \in \calX}$, define
\[
h(\bfp) = \sum_{x \in \calX} -p_x \cdot \ln p_x = \sum_{x \in \calX} \eta(p_x) \; .
\]

%

For $A = \mysett{y_1,y_2,\ldots,y_t} \subseteq \calY$, define the quantity $\Delta(A)$ as the decrease in mutual information resulting from merging all symbols in $A$ into a single symbol in $Q$. Namely, define 
\begin{equation}
\label{eq:DeltaDefinition}
\Delta(A) \triangleq \pi \left(  h\left[\sum_{j=1}^t \theta_j \bfp^{(j)}\right] -\left( \sum_{j=1}^t \theta_j h\big[\bfp^{(j)}\big] \right) \right) \; ,
\end{equation}
where
\begin{equation}
\label{eq:piAndThetaDefined}
\pi = \sum_{y \in A} P_Y(y) \; , \quad
\theta_j = P_Y(y_j)/\pi \; , 
\end{equation}
and
\begin{equation}
\label{eq:bfpDefined}
\bfp^{(j)} = \probvectorr{P(X=x|Y=y_j)}_{x \in \calX} \; .
\end{equation}

The following claim is easily derived.

\begin{claim}
\label{claim:costExact}
Let $W$, $Q$, $P_X$, $L$, and $(A_i)_{i=1}^L$  be as in Claim~\ref{cl:optimalQuantization}. Then,
\begin{equation}
\label{eq:capacityDifferenceInTermsOfDelta}
I(W) - I(Q) = \sum_{i=1}^L \Delta(A_i) \; .
\end{equation}
\end{claim}


Although the drop in mutual information is easily described, we were not able to analyze and manipulate it directly. We now aim for a bound which is more amenable to analysis. As mentioned, by the concavity of $h$ and Jensen's inequality, we deduce that $\Delta(A_i) \geq 0$. Namely, data processing reduces mutual information. We will shortly make use of the fact that $h$ is strongly concave in order to derive a sharper lower bound. To that end, we now state H\"older's defect formula \cite{Holder:1889p} (see \cite[Page 94]{Steele:04b} for an accessible reference).

As is customary, we will phrase H\"older's defect formula for $\cup$-convex functions, although we will later apply it to $h$ which is $\cap$-concave. We remind the reader that for twice differentiable $\cup$-convex functions, $f \colon D \to \reals$, $D \subseteq \reals^n$, the Hessian of $f$, denoted
\[
\nabla^2 f(\alpha) = \left(\frac{\partial^2 f(\alpha)}{\partial \alpha_i \partial \alpha_j}\right)_{i,j} \; ,
\]
is positive semidefinite on the interior of $D$ \cite[page 71]{BoydVandenberghe:04b}. We denote the smallest eigenvalue of $\nabla^2 f(\alpha)$ by $\lambdamin(\nabla^2 f(\alpha))$. 
\begin{lemm}
\label{lemm:holder}
Let $f(\alpha) \colon D \to \reals$ be a twice differentiable convex function defined over a convex domain $D \subseteq \reals^n$. Let $m \geq 0$ be such that for all $\alpha$ in the interior of $D$,
\[
m \leq \lambdamin(\nabla^2 f(\alpha))
\]
Fix $(\alpha_j)_{j=1}^t \in D$ and let $(\theta_j)_{j=1}^t$ be non-negative coefficients summing to $1$. Denote
\[
\overline{\alpha} = \sum_{j=1}^t \theta_j \alpha_j
\]
and
\[
\delta^2 = \sum_{j=1}^t \theta_j \normts{\alpha_j - \overline{\alpha}} = \frac{1}{2} \sum_{j=1}^t \sum_{k=1}^t \theta_j \theta_k \normts{\alpha_j - \alpha_k}
\]
Then, 
\[
\sum_{j=1}^t \theta_j f[\alpha_j] - f[\sum_j \theta_j \alpha_j] \geq \frac{1}{2} m\delta^2 \; .
\]
\end{lemm}

\begin{IEEEproof}
Let $\Lambda$ be a diagonal matrix having all entries equal to $m$. By definition of $m$, we have that the function $g(\alpha) = f(\alpha) - \frac{1}{2} \alpha^T \Lambda \alpha$ is positive semidefinite for all $\alpha \in D$. Thus, by Jensen's inequality,  
\[
\sum_i \theta_i g[\alpha_i] - g[\sum_i \theta_i \alpha_i] \geq 0 \; .
\]
Replacing $g(\alpha)$ in the above expression by  $f(\alpha) - \frac{m}{2} \alpha^T \alpha$ and rearranging yields the required result.
\end{IEEEproof}

We now apply H\"older's inequality in order to bound $\Delta(A)$. For $A = \mysett{y_1,y_2,\ldots,y_t} \subseteq \calY$, define  
\begin{condmultline}
\label{eq:Deltatilde}
\Deltatilde(A) \triangleq \frac{\pi}{2} \sum_{j=1}^t \theta_j \normts{\bfp^{(j)} - \bar{\bfp}} \\
= \frac{\pi}{4} \sum_{j=1}^t \sum_{k=1}^t \theta_j \theta_k \normts{\bfp^{(j)} - \bfp^{(k)}} \; ,
\end{condmultline}
where $\pi$ and $\theta_j$ are as in (\ref{eq:piAndThetaDefined}), $\bfp^{(j)}$ is as defined in (\ref{eq:bfpDefined}), and 
\[
\bar\bfp = \sum_{j=1}^t \theta_j \bfp(j) \; .
\]
The following is a simple corollary of Lemma~\ref{lemm:holder}
\begin{coro}
\label{coro:degradingCostBoundedByDeltatildesum}
Let $W$, $Q$, $P_X$, $L$, and $(A_i)_{i=1}^L$  be as in Claim~\ref{cl:optimalQuantization}. Then, for all $1 \leq i \leq L$,
\begin{equation}
\label{eq:DeltaDeltatildeBound}
\Delta(A_i) \geq \Deltatilde(A_i) \; .
\end{equation}
Thus,
\begin{equation}
\label{eq:degradingCostBoundedByDeltatildesum}
I(W) - I(Q) \geq \sum_{i=1}^L \Deltatilde(A_i) \; .
\end{equation}
\end{coro}
\begin{IEEEproof}
The second inequality follows from the first inequality and (\ref{eq:capacityDifferenceInTermsOfDelta}). We now prove the first inequality.
Let $D = [0,1]^n$, the set of vectors of length $n$ having each entry between $0$ and $1$. Since the second derivative of $\eta$ is $\eta''(p) = -1/p$, we conclude $\lambdamin(-h(\bfp)) \geq 1$ for all $\bfp$ in the interior $(0,1)^n$. That is, we take $m=1$ in Lemma~\ref{lemm:holder}. Since $h$ is continuous on $D$,  our result follows by Lemma~\ref{lemm:holder} and standard limiting arguments.
\end{IEEEproof}

\section{Bounding the degrading cost}
\label{sec:boundingTheDegradingCost}
We now turn to bounding the degrading cost. As a first step, we define a channel $\Whard$ for which we will prove a lower bound on the cost of degrading.

\subsection{The channel $\Whard$}
For a specified integer $M \geq 1$, we now define the channel $\Whard = \Whard_M$, where $\Whard \colon \calX \to \calY$. The input alphabet is $\calX = \mysett{1,2,\ldots,q}$, of size $\mysizee{\calX} = q$. The output alphabet consists of vectors of length $q$ with integer entries, defined as follows:
\begin{condmultline}
\label{eq:calYdef}
\calY = \Big\{\, \probvectorm{j_1,j_2,\ldots,j_q} :  \\
j_1,j_2,\ldots,j_q \geq 0 \; , \quad \sum_{x=1}^q j_x = M\, \Big\} \; .
\end{condmultline}
The channel transition probabilities are given by 
\[
\Whard(\probvectorm{j_1,j_2,\ldots,j_q} | x ) = \frac{q \cdot j_x}{M \binom{M+q-1}{q-1}} \; .
\] 

\begin{lemm}
\label{lemm:WhardValid}
The above defined $\Whard$ is a valid channel with output alphabet size
\begin{equation}
\label{eq:sizeOfWhardNonbinary}
\channelSize{\Whard} = \binom{M+q-1}{q-1} \; .
\end{equation}
\end{lemm}
\begin{IEEEproof}
The binomial expression for the output alphabet size follows by noting that we are essentially dealing with an instance of ``combinations with repetitions'' \cite[Page 15]{Stanley:97b}. Obviously, the probabilities are non-negative. It remains to show that for all $x \in \calX$,  
\[
\sum_{\probvectorm{j_1,j_2,\ldots,j_q} \in \calY} \quad \frac{q \cdot j_x}{M \binom{M+q-1}{q-1}} = 1 \; .
\]
Since the above is independent of $x$, we can equivalently show that
\[
\sum_{\probvectorm{j_1,j_2,\ldots,j_q} \in \calY} \quad \frac{q \cdot (j_1 + j_2 + \cdots + j_q)}{M \binom{M+q-1}{q-1}} = q \; .
\]
By the definition of $\calY$ in (\ref{eq:calYdef}), the denominator above equals $q \cdot M$. Since we have already proved (\ref{eq:sizeOfWhardNonbinary}), the result follows.
\end{IEEEproof}

Recall the definition of symmetry in \cite[page 94]{Gallager:68b}: Let $W : \calX \to \calY$ be a channel. Define the probability matrix associated with $W$ as a matrix with rows indexed by $\calX$ and columns by $\calY$ such that entry $(x,y) \in \calX \times \calY$ equals $W(y|x)$. The channel $W$ is symmetric if the output alphabet can be partitioned into sets, and the following holds: for each set, the corresponding submatrix  is such that every row is a permutation of the first row and every column is a permutation of the first column.
\begin{lemm}
The above defined $\Whard$ is a symmetric channel.
\end{lemm}
\begin{IEEEproof}
Define the partition so that two output letters, $\probvectorm{j_1,j_2,\ldots,j_q}$ and $\probvectorm{j'_1,j'_2,\ldots,j'_q}$, are in the same set if there exists a permutation $\pi : \calX \to \calX$ such that $j_x = j'_{\pi(x)}$, for all $x \in \calX$.
\end{IEEEproof}

Since $\Whard$ is symmetric, it follows from \cite[Theorem 4.5.2]{Gallager:68b} that the capacity achieving distribution is the uniform distribution. Thus, we take the corresponding input distribution as uniform. Namely, for all $x \in \calX$,
\[
P(X=x) = \frac{1}{q} \; .
\]
As a result, all output letters are equally likely (the proof is similar to that of Lemma~\ref{lemm:WhardValid}).

Denote the vector of  a posteriori probabilities corresponding to $\probvectorm{j_1,j_2,\ldots,j_q}$ as
\[
\bfp(j_1,j_2,\ldots,j_q) = \probvectorr{\; P(X = x | Y = \probvectorm{j_1,j_2,\ldots,j_q}) \;}_{x=1}^q \; .
\]
A short calculation gives
\begin{equation}
\label{eq:bfpNonbinary}
\bfp(j_1,j_2,\ldots,j_q) = \probvector{ \frac{j_1}{M}, \frac{j_2}{M},\ldots,\frac{j_q}{M} }  \; .
\end{equation}
In light of the above, let us define the shorthand
\[
\probvectorm{j_1,j_2,\ldots,j_q} \triangleq \probvectorr{j_1/M,j_2/M,\ldots,j_q/M} \; .
\]
With this shorthand in place, the label of each output letter $\probvectorm{j_1,j_2,\ldots,j_q} \in \calY$ is the corresponding a posteriori probability vector $\bfp(j_1,j_2,\ldots,j_q)$. Thus, we gain a simple expression for $\Deltatilde(A)$. Namely, for $A \subseteq \calY$,
\[
\Deltatilde(A) = \frac{1}{2 \binom{M+q-1}{q-1}} \sum_{\bfp \in A} \normts{\bfp - \bar{\bfp}} \; , \quad \bar{\bfp} = \sum_{\bfp \in A} \frac{1}{\mysizee{A}} \bfp \; .
\]

We remark in passing that as $M \to \infty$, $\Whard$ ``converges'' to the channel $\calW_q \colon \calX \to \calX \times [0,1]^q$ which we now define. Given an input $x$, the channel picks $\varphi_1,\varphi_2,\ldots,\varphi_q$ as follows:  $\varphi_1,\varphi_2,\ldots,\varphi_{q-1}$ are picked according to the Dirichlet distribution $D(1,1,\ldots,1)$, while $\varphi_q$ is set to $1 - \sum_{x=1}^{q-1} \varphi_x$. That is, $(\varphi_1,\varphi_2,\ldots,\varphi_q)$ is chosen uniformly from all possible probability vectors of length $q$. Then, the input $x$ is transformed into $x+i$ (with a modulo operation where appropriate\footnote{To be precise, $x$ is transformed into $1 + (x -1 + i \mod q)$.}) with probability $\varphi_i$. The transformed symbol along with the vector $(\varphi_1,\varphi_2,\ldots,\varphi_q)$ is the output of the channel.

\subsection{Optimizing $A'$}
Our aim is to find a lower bound on $\Deltatilde(A)$, where $A \subseteq \calY$ is constrained to have a size $\mysizee{A} = t$. Recalling  (\ref{eq:bfpNonbinary}), note that all output letters $\bfp = \probvectorr{p_x}_{x=1}^q \in \calY$ must satisfy the following three properties.
\begin{enumerate}
\item All entries $p_x$ are of the form $j_x/M$, where $j_x$ is an integer.
\item \label{it:allPxEntriesSumToOne} All entries $p_x$ sum to $1$.
\item \label{it:allPxEntriesNonNegative} All entries $p_x$ are non-negative.
\end{enumerate}
Since all entries must sum to $1$ by property \ref{it:allPxEntriesSumToOne}, entry $p_{q}$ is redundant. Thus, for a given $\bfp \in \calY$, denote by $\bfp'$ the first $q-1$ coordinates of $\bfp$. Let $A'$ be the set one gets by applying this puncturing operation to each element of $A$. Denote 
\begin{equation}
\label{eq:DeltatildeShortenedP}
\Deltatilde(A') \triangleq \frac{1}{2 \binom{M+q-1}{q-1}} \sum_{\bfp' \in A'} \normts{\bfp' - \bar{\bfp}'} \; ,
\end{equation}
One easily shows that
\begin{equation}
\label{eq:deltaTildeAprimeVersusA}
\Deltatilde(A') \leq \Deltatilde(A) \; ,
\end{equation}
thus a lower bound on $\Deltatilde(A')$ is also a lower bound on $\Deltatilde(A)$.

In order to find a lower bound on $\Deltatilde(A')$ we relax constraint \ref{it:allPxEntriesNonNegative} above. Namely, a set $A'$ with elements $\bfp'$ will henceforth mean a set for which each element $\bfp' = \probvectorr{p_x}_{x=1}^{q-1}$ has entries of the form $p_x = j /M$, and each such entry is \emph{not} required to be non-negative. Our revised aim is to find a lower bound on $\Deltatilde(A')$ where $A'$ holds elements as just defined and is constrained to have size $t$. The simplification enables us to give a characterization of the optimal $A'$. Informally, a sphere, up to irregularities on the boundary.

\begin{lemm}
Let $t > 0$ be a given integer. Let $A'$ be the set of size $\mysizee{A'} = t$ for which $\Deltatilde(A')$ is minimized. Denote by $\bar{\bfp}'$ the mean of all elements of $A'$. Then, $A'$ has a critical radius $r$: all $\bfp'$ for which $\normts{\bfp'-\bar{\bfp}'} < r^2$ are in $A'$ and all $\bfp'$ for which $\normts{\bfp'-\bar{\bfp}'} > r^2$ are not in $A'$.
\end{lemm}

\begin{IEEEproof}
We start by considering a general $A'$. Suppose $\bfp'(1) \in A'$ is such that $r^2 = \normts{\bfp'(1) - \bar{\bfp}'}$. Next, suppose that there is a $\bfp'(2) \not\in A'$ such that $\normts{\bfp'(2) - \bar{\bfp}'} < r^2$. Then, for
\[
B' = A' \cup \mysett{\bfp'(2)} \setminus \mysett{\bfp'(1)} \; , \quad
\Deltatilde(B') < \Deltatilde(A') \; .
\]
To see this, first note that
\begin{equation}
\label{eq:BversusAInSums}
\sum_{\bfp' \in B'} \normts{\bfp' - \bar{\bfp}'} < 
\sum_{\bfp' \in A'} \normts{\bfp' - \bar{\bfp}'} \; .
\end{equation}
Next, note that the RHS of (\ref{eq:BversusAInSums}) is $\Deltatilde(A')$, but the LHS is \emph{not} $\Deltatilde(B')$. Namely, $\bar{\bfp}'$ is the mean of the vectors in $A'$ but is not the mean of the vectors in $B'$. However, $\sum_{\bfp' \in B'} \normts{\bfp' - \bfu'}$ is minimized for $\bfu'$ equal to the mean of the vectors in $B'$ (to see this, differentiate the sum with respect to every coordinate of $\bfu'$). Thus, the LHS of (\ref{eq:BversusAInSums}) is at least $\Deltatilde(B')$ while the RHS equals $\Deltatilde(A')$.

The operation of transforming $A'$ into $B'$ as above can be applied repeatedly, and must terminate after a finite number of steps. To see this, note that the sum $\sum_{\bfp' \in A'} \normts{\bfp' - \bar{\bfp}'}$ is constantly decreasing, and so is upper bounded by the initial sum. Therefore, one can bound the maximum distance between any two points in $A'$. Since the sum is invariant to translations, we can always translate $A'$ such that its members are contained in a suitably large hypercube (the translation will preserve the $1/M$ grid property). The number of ways to distribute $|A'|$ grid points inside the hypercube is finite. Since the sum is strictly decreasing and non-negative, the number of steps is finite. The ultimate termination implies a critical $r$ as well as the existence of an optimal $A'$. 
\end{IEEEproof}

Recall that a sphere of radius $r$ in $\reals^{q-1}$ has volume $\sigma_{q-1}r^{q-1}$, where $\sigma_{q-1}$ is a well known constant \cite[Page 411]{Apostol:67.2b}. Given a set $A'$, we define the volume of $A'$ as
\[
\Vol(A') \triangleq \frac{\mysizee{A'}}{M^{q-1}} \; .
\]
For optimal $A'$ as above, the following lemma approximates $\Vol(A')$ by the volume of a corresponding sphere.

\begin{lemm}
\label{lemm:volumeOfAPrime}
Let $A'$ be a set of size $t$ for which $\Deltatilde(A')$ is minimized. Let the critical radius be $r$ and assume that $r \leq 4$. Then,
\[
\Vol(A') = \sigma_{q-1}r^{q-1} + \epsilon_{q-1}(t) \; .
\]
The error term $\epsilon_{q-1}(t)$ is bounded from both above and below by functions of $M$ alone (\emph{not} of $t$) that are $o(1)$ (decay to $0$ as $M \to \infty$).
\end{lemm}

\begin{IEEEproof}
Let $\delta \colon \reals^{q-1} \to \mysett{0,1}$ be the indicator function of a sphere with radius $r$ centered at $\bar{\bfp}'$. That is,
\[
\delta(\bfp') = 
\begin{cases}
1 & \normts{\bfp' - \bar{\bfp}'} \leq r^2 \\
0 & \mbox{otherwise} \; .
\end{cases}
\]
Note that 1) $\delta$ is a bounded function and 2) the measure of points for which $\delta$ is not continuous is zero (the boundary of a sphere has no volume). Thus, $\delta$ is Riemann integrable \cite[Theorem 14.5]{Apostol:74b}.

Consider the set $\Psi'$ which is $[-4r,4r]^{q-1}$ shifted by $\bar{\bfp}'$. Since $\Psi'$ contains the above sphere, the integral of $\delta$ over $\Psi'$ must equal $\sigma_{q-1} r^{q-1}$. We now show a specific Riemann sum \cite[Definition 14.2]{Apostol:74b} which must converge to this integral. Consider a partition of $\Psi'$ into cubes of side length $1/M$, where each cube center is of the form $\probvectorr{j_1/M,j_2/M,\ldots,j_{q-1}/M}$ and the $j_x$ are integers (the fact that cubes at the edge of $\Psi'$ are of volume less than $1/M^{q-1}$ is immaterial).  Define $[\bfp' \in A']$ as $1$ if the condition $\bfp' \in A'$ holds and $0$ otherwise. We claim that the following is a Riemann sum of $\delta$ over $\Psi'$ with respect to the above partition. 
\[
\sum_{ \bfp'  = \probvectorr{j_1/M,j_2/M,\ldots,j_{q-1}/M} \in \Psi' } \frac{1}{M^{q-1}} [ \bfp' \in A' ]
\]
To see this, recall that $A'$ has critical radius $r$.

The absolute value of the difference between the above sum and $\sigma_{q-1} r^{q-1}$ can be upper bounded by the number of cubes that straddle the sphere times their volume $1/M^{q-1}$ (any finer partition will only affect these cubes). Since $r \leq 4$, this quantity must go to zero as $M$ grows, no matter how we let $r$ depend on $M$.
\end{IEEEproof}

\begin{lemm}
\label{lemm:DeltatildeInTermsOfRadius}
Let $A'$ be a set of size $t$ for which $\Deltatilde(A')$ is minimized. Let the critical radius be $r$ and assume that $r \leq 4$. Then,
\[
\Deltatilde(A') = \frac{(q-1) \cdot (q-1)!}{2(q+1)} \sigma_{q-1} r^{q+1} + \epsilon_{q-1}(t) \; .
\]
The error term $\epsilon_{q-1}(t)$ is bounded from both above and below by functions of $M$ alone (\emph{not} of $t$) that are $o(1)$ (decay to $0$ as $M \to \infty$).
\end{lemm}

\begin{IEEEproof}
Let the sphere indicator function $\delta$ and the bounding set $\Psi'$ be as in the proof of Lemma~\ref{lemm:volumeOfAPrime}. Consider the sum
\begin{equation}
\label{eq:riemannOfDelta}
\sum_{ \bfp'  = \probvectorr{j_1/M,j_2/M,\ldots,j_{q-1}/M} \in \Psi' } \frac{1}{M^{q-1}} \normts{\bfp' - \bar{\bfp}'} \cdot [ \bfp' \in A' ] \; .
\end{equation}
On the one hand, by (\ref{eq:DeltatildeShortenedP}), this sum is simply
\begin{equation}
\label{eq:riemannOfDeltaSimplification}
\frac{2 \binom{M+q-1}{q-1}}{M^{q-1}} \Deltatilde(A') \; .
\end{equation}
On the other hand, (\ref{eq:riemannOfDelta}) is the Riemann sum corresponding to the integral
\[
\int_{\Psi'} \normts{\bfp' - \bar{\bfp}'} [\bfp' \in A'] \, d\bfp' \; ,
\]
with respect to the same partition as was used in the proof of Lemma~\ref{lemm:volumeOfAPrime}. As before, the sum must converge to the integral, and the convergence rate can be shown to be bounded by expressions which are not a function of $t$.

All that remains is to calculate the integral. Denote by $\sphere_{q-1}(r) \subseteq \reals^{q-1}$ the sphere centered at the origin with radius $r$. After translating $\bar{\bfp}'$ to the origin, the integral becomes
\begin{condmultline}
\label{eq:intOfraduisSquared}
\int_{\sphere_{q-1}(r)} \left( x_1^2 + x_2^2 + \cdots + x_{q-1}^2 \right) \;  dx_1 dx_2 \cdots dx_{q-1} \\
= \frac{\sigma_{q-1}\cdot (q-1) \cdot r^{q+1}}{q+1} \; ,
\end{condmultline}
where the RHS is derived as follows. After converting the integral to generalized spherical coordinates
\begin{IEEEeqnarray*}{rcl}
x_1 &=& r \cos(\theta_1) \; , \\
x_2 &=& r \sin(\theta_1) \cos(\theta_2) \; , \\
& \vdots& \\
x_{q-2} &=& r \sin(\theta_1) \sin(\theta_2) \cdots \sin(\theta_{q-2}) \cos(\theta_{q-1})\; , \\
x_{q-1} &=& r \sin(\theta_1) \sin(\theta_2) \cdots \sin(\theta_{q-2}) \sin(\theta_{q-1}) \; ,
\end{IEEEeqnarray*}

we get an integrand that is $r^2$ times the integrand we would have gotten had the original integrand been $1$ (this follows by applying the identity $\sin^2 \theta + \cos^2 \theta = 1$ repeatedly). We know that had that been the case, the integral would have equaled $\sigma_{q-1} r^{q-1}$.

Since (\ref{eq:intOfraduisSquared}) must equal the limit of (\ref{eq:riemannOfDeltaSimplification}), and since the fraction in (\ref{eq:riemannOfDeltaSimplification}) converges to $2/(q-1)!$, the claim follows.
\end{IEEEproof}

As a corollary to the above three lemmas, we have the following result. The important point to note is that the RHS is convex in $\Vol(A')$.
\begin{coro}
\label{coro:boundOnDeltatilde}
Let $t > 0$ be a given integer. Let $A'$ be a set of size $t$ and assume that
\begin{equation}
\label{eq:bfpprimeCondition}
\max_{\bfp' \in A'} \normts{\bfp' - \bar{\bfp}'} \leq 2 \; .
\end{equation}
Then,
\begin{equation}
\label{eq:boundOnDeltaTildeViaVolume}
\Deltatilde(A') \geq \frac{(q-1)\cdot (q-1)!}{2(q+1)\cdot (\sigma_{q-1})^{\frac{2}{q-1}}} \cdot  \Vol(A')^{\frac{q+1}{q-1}} + o(1) \; ,
\end{equation}
where the $o(1)$ is a function of $M$ alone and goes to $0$ as $M \to \infty$.
\end{coro}
\begin{IEEEproof}
Let $B'$ be the set of size $t$ for which $\Deltatilde(B')$ is minimized. The proof centers on showing that the critical radius of $B'$ is at most $4$. All else follows directly from Lemmas~\ref{lemm:volumeOfAPrime} and \ref{lemm:DeltatildeInTermsOfRadius}. Assume to the contrary that the critical radius of $B'$ is greater than $4$. Thus, up to translation, $A'$ is a subset of $B'$. But this implies that $\Deltatilde(A') < \Deltatilde(B')$, a contradiction.
\end{IEEEproof}

\subsection{Bounding $\degradingCost(q,L)$}
We are now in a position to prove Theorem~\ref{theo:generalBound}. Recall that $A_i$ is the set of output letters in $\calY$ which get mapped to the letter $z_i \in \calZ$. Also, recall that $A'_i$ is simply $A_i$ with the last entry dropped from each vector.
\begin{IEEEproof}[Proof of Theorem~\ref{theo:generalBound}]
By combining (\ref{eq:degradingCostDefinition}), (\ref{eq:degradingCostBoundedByDeltatildesum}), (\ref{eq:deltaTildeAprimeVersusA}), and (\ref{eq:boundOnDeltaTildeViaVolume}), we have that as long as condition (\ref{eq:bfpprimeCondition}) holds for all $A'_i$, $1 \leq i \leq L$, the degrading cost $\degradingCost(q,L)$ is at least
\begin{equation}
\label{eq:boundOnDegradingCostWithAprimes}
\frac{(q-1)\cdot (q-1)!}{2(q+1)\cdot (\sigma_{q-1})^{\frac{2}{q-1}}} \sum_{i=1}^L  \Vol(A_i')^{\frac{q+1}{q-1}} + o(1) \; .
\end{equation}
Recalling that the elements of $A$ are probability vectors, we deduce that condition (\ref{eq:bfpprimeCondition}) must indeed hold. Indeed,
\[
\normts{\bfp' - \bar{\bfp}'} \leq \normt{\bfp' - \bar{\bfp}'} \leq \normt{\bfp'} + \normt{\bar{\bfp}'} \leq 2 \; .
\]
The first inequality follows from the fact that $p^2$ is less than $p$ for $0 \leq p \leq 1$. The second inequality is the triangle inequality. The third inequality follows from the same reasons as the first.

 Next, recall that $\Vol(A_i') = \Vol(A_i)$, and thus
\begin{equation}
\label{eq:sumVolumesConstraint}
\sum_{i=1}^L \Vol(A_i') = \frac{\channelSize{\Whard}}{M^{q-1}} = \frac{\binom{M+q-1}{q-1}}{M^{q-1}} \; . 
\end{equation}
Note that the RHS converges to $1/(q-1)!$ as $M \to \infty$.
By convexity, we have that if we are constrained by (\ref{eq:sumVolumesConstraint}), then the sum in (\ref{eq:boundOnDegradingCostWithAprimes}) is lower bounded by setting all $\Vol(A_i')$ equal to the RHS of (\ref{eq:sumVolumesConstraint}) divided by $L$. Thus, after taking $M \to \infty$, we get (\ref{eq:generalBound}).
\end{IEEEproof}
\textbf{Acknowledgments}: The author thanks Eren \c{S}a\c{s}o\u{g}lu and Igal Sason for their feedback.

\twobibs{
\bibliographystyle{IEEEtran}
\bibliography{\bibfilePath}

\begin{thebibliography}{10}
\providecommand{\url}[1]{#1}
\csname url@samestyle\endcsname
\providecommand{\newblock}{\relax}
\providecommand{\bibinfo}[2]{#2}
\providecommand{\BIBentrySTDinterwordspacing}{\spaceskip=0pt\relax}
\providecommand{\BIBentryALTinterwordstretchfactor}{4}
\providecommand{\BIBentryALTinterwordspacing}{\spaceskip=\fontdimen2\font plus
\BIBentryALTinterwordstretchfactor\fontdimen3\font minus
  \fontdimen4\font\relax}
\providecommand{\BIBforeignlanguage}[2]{{%
\expandafter\ifx\csname l@#1\endcsname\relax
\typeout{** WARNING: IEEEtran.bst: No hyphenation pattern has been}%
\typeout{** loaded for the language `#1'. Using the pattern for}%
\typeout{** the default language instead.}%
\else
\language=\csname l@#1\endcsname
\fi
#2}}
\providecommand{\BIBdecl}{\relax}
\BIBdecl

\bibitem{Arikan:09p}
\bibstar{isf2013}E. Ar{\i}kan, ``Channel polarization: A method for
  constructing capacity-achieving codes for symmetric binary-input memoryless
  channels,'' \emph{IEEE Trans. Inform. Theory}, vol.~55, pp. 3051--3073, 2009.

\bibitem{STA:09a}
\bibstar{isf2013}E. \c{S}a\c{s}o\u{g}lu, E.~Telatar, and E.~Ar{\i}kan,
  ``Polarization for arbitrary discrete memoryless channels,''
  \emph{\textup{\texttt{arXiv:0908.0302v1}}}, 2009.

\bibitem{Gallager:68b}
R.~G. Gallager, \emph{Information Theory and Reliable Communications}.\hskip
  1em plus 0.5em minus 0.4em\relax New York: John Wiley, 1968.

\bibitem{MoriTanaka:10c}
R.~Mori and T.~Tanaka, ``Non-binary polar codes using reed-solomon codes and
  algebraic geometry codes,'' in \emph{Proc. IEEE Inform. Theory Workshop
  (ITW'2010)}, Dublin, Ireland, 2010, pp. 1--5.

\bibitem{MoriTanaka:14p}
------, ``Source and channel polarization over finite fields and reed–solomon
  matrices,'' vol.~60, 2014, pp. 2720--2736.

\bibitem{PresmanShapiraLitsyn:11c}
N.~Presman, O.~Shapira, and S.~Litsyn, ``Polar codes with mixed kernels,'' in
  \emph{Proc. IEEE Int'l Symp. Inform. Theory (ISIT'2011)}, Saint Petersburg,
  Russia, 2011, pp. 6--10.

\bibitem{Sasoglu:12c}
E.~\c{S}a\c{s}o\u{g}lu, ``Polar codes for discrete alphabets,'' in \emph{Proc.
  IEEE Int'l Symp. Inform. Theory (ISIT'2012)}, Cambridge, Massachusetts, 2012,
  pp. 2137--21\,141.

\bibitem{RichardsonUrbanke:08b}
T.~Richardson and R.~Urbanke, \emph{Modern Coding Theory}.\hskip 1em plus 0.5em
  minus 0.4em\relax Cambridge, UK: Cambridge University Press, 2008.

\bibitem{MoriTanaka:09c}
R.~Mori and T.~Tanaka, ``Performance and construction of polar codes on
  symmetric binary-input memoryless channels,'' in \emph{Proc. IEEE Int'l Symp.
  Inform. Theory (ISIT'2009)}, Seoul, South Korea, 2009, pp. 1496--1500.

\bibitem{TalVardy:13p}
I.~Tal and A.~Vardy, ``How to construct polar codes,'' \emph{IEEE Trans.
  Inform. Theory}, vol.~59, pp. 6562--6582, 2013.

\bibitem{PHTT:11c}
\bibstar{isf2013}R. Pedarsani, S.~H. Hassani, I.~Tal, and E.~Telatar, ``On the
  construction of polar codes,'' in \emph{Proc. IEEE Int'l Symp. Inform. Theory
  (ISIT'2011)}, Saint Petersburg, Russia, 2011, pp. 11--15.

\bibitem{KurkoskiYagi:11a}
B.~M. Kurkoski and H.~Yagi, ``Quantization of binary-input discrete memoryless
  channels, with applications to {LDPC} decoding,''
  \emph{\textup{\texttt{arXiv:11107.5637v1}}}, 2011.

\bibitem{TalSharovVardy:12c}
\bibstar{isf2013}I. Tal, A.~Sharov, and A.~Vardy, ``Constructing polar codes
  for non-binary alphabets and {MAC}s,'' in \emph{Proc. IEEE Int'l Symp.
  Inform. Theory (ISIT'2012)}, Cambridge, Massachusetts, 2012, pp. 2132--2136.

\bibitem{PeregTal:15a}
U.~Pereg and I.~Tal, ``Channel upgradation for non-binary input alphabets and
  {MACs},'' \emph{\textup{\texttt{arXiv:1308.5793v3}}}, 2015.

\bibitem{GhayooriGulliver:12a}
A.~Ghayoori and T.~A. Gulliver, ``Upgraded approximation of non-binary
  alphabets for polar code construction,''
  \emph{\textup{\texttt{arXiv:1304.1790v3}}}, 2013.

\bibitem{GuruswamiVelingker:14a}
V.~Guruswami and A.~Velingker, ``An entropy sumset inequality and polynomially
  fast convergence to shannon capacity over all alphabets,''
  \emph{\textup{\texttt{arXiv:1411.6993}}}, 2014.

\bibitem{Holder:1889p}
O.~H{\"o}lder, ``{\"U}ber einen mittelwerthssatz,'' \emph{Nachr. Akad. Wiss.
  G{\"o}ttingen Math.-Phys. Kl.}, pp. 38--47, 1889.

\bibitem{Steele:04b}
\emph{The Cauchy--Schwarz Master Class}.\hskip 1em plus 0.5em minus 0.4em\relax
  Cambridge, UK: Cambridge University Press, 2004.

\bibitem{BoydVandenberghe:04b}
S.~Boyd and L.~Vandenberghe, \emph{Convex Optimization}.\hskip 1em plus 0.5em
  minus 0.4em\relax Cambridge, UK: Cambridge University Press, 2004.

\bibitem{Stanley:97b}
R.~P. Stanley, \emph{Enumerative Combinatorics}.\hskip 1em plus 0.5em minus
  0.4em\relax Cambridge, UK: Cambridge University Press, 1997, vol.~1.

\bibitem{Apostol:67.2b}
T.~M. Apostol, \emph{Calculus}, 2nd~ed.\hskip 1em plus 0.5em minus 0.4em\relax
  Wiley, 1967, vol.~2.

\bibitem{Apostol:74b}
------, \emph{Mathematical Analysis}, 2nd~ed.\hskip 1em plus 0.5em minus
  0.4em\relax Reading, Massachusetts: Addison-Wesley, 1974.

\end{thebibliography}
}
{
\ifdefined\bibstar\else\newcommand{\bibstar}[1]{}\fi

}

\end{document}